\documentclass[a4paper,11pt]{article}
\usepackage{pos}
\pdfoutput=1

\usepackage{graphicx}
\usepackage{xcolor}
\usepackage{booktabs}
\usepackage{caption}
\usepackage{xspace}

\newcommand\qpfour{QPACE~4\xspace}

\begin{document}

\title{Grid on \qpfour}

\ShortTitle{Grid on \qpfour}

\author{Peter Georg}
\author*{Nils Meyer}
\author{Stefan Solbrig}
\author{Tilo Wettig}

\affiliation{Department of Physics, University of Regensburg, 93040 Regensburg, Germany}

\emailAdd{nils.meyer@ur.de}

\abstract{In 2020 we deployed \qpfour, which features 64 Fujitsu A64FX model
FX700 processors interconnected by InfiniBand EDR.  \qpfour runs an open-source
software stack.  For Lattice QCD simulations we ported the Grid LQCD framework to
support the ARM Scalable Vector Extension (SVE).  In this contribution we
discuss our SVE port of Grid, the status of SVE compilers and the performance
of Grid.  We also present the benefits of an alternative data layout of complex
numbers for the Domain Wall operator.}

\FullConference{%
 The 38th International Symposium on Lattice Field Theory, LATTICE2021
  26th-30th July, 2021
  Zoom/Gather@Massachusetts Institute of Technology
}

\maketitle

\section{Introduction}

Lattice QCD has traditionally been a developer and early adopter of new
high-performance computing (HPC) resources.  A very promising new instruction
set architecture is Arm's Scalable Vector Extension (SVE), which has been
implemented in hardware for the first time in the A64FX processor developed by
RIKEN R-CCS and Fujitsu.  This processor powers Japan's flagship Fugaku
supercomputer and is also used in the \qpfour cluster at the University of
Regensburg.

Porting and optimizing Lattice QCD code for a new hardware architecture has been
a difficult and time-consuming task in the past.  The Grid Lattice QCD framework
\cite{Boyle:2015tjk} makes this task much easier by isolating the details of the
hardware in a few files at the lowest layer of the software stack.  Our port of
Grid to SVE is described in \cite{Meyer:cluster18, Meyer:lattice18, Meyer:aplat20}.
In this contribution we study the performance of this port on \qpfour, including
compiler performance and a different choice of the data layout for complex
numbers compared to Grid.

\begin{figure}[b]
\begin{minipage}[t]{0.47\linewidth}
        \begin{center}
            \includegraphics[width=0.99\linewidth]{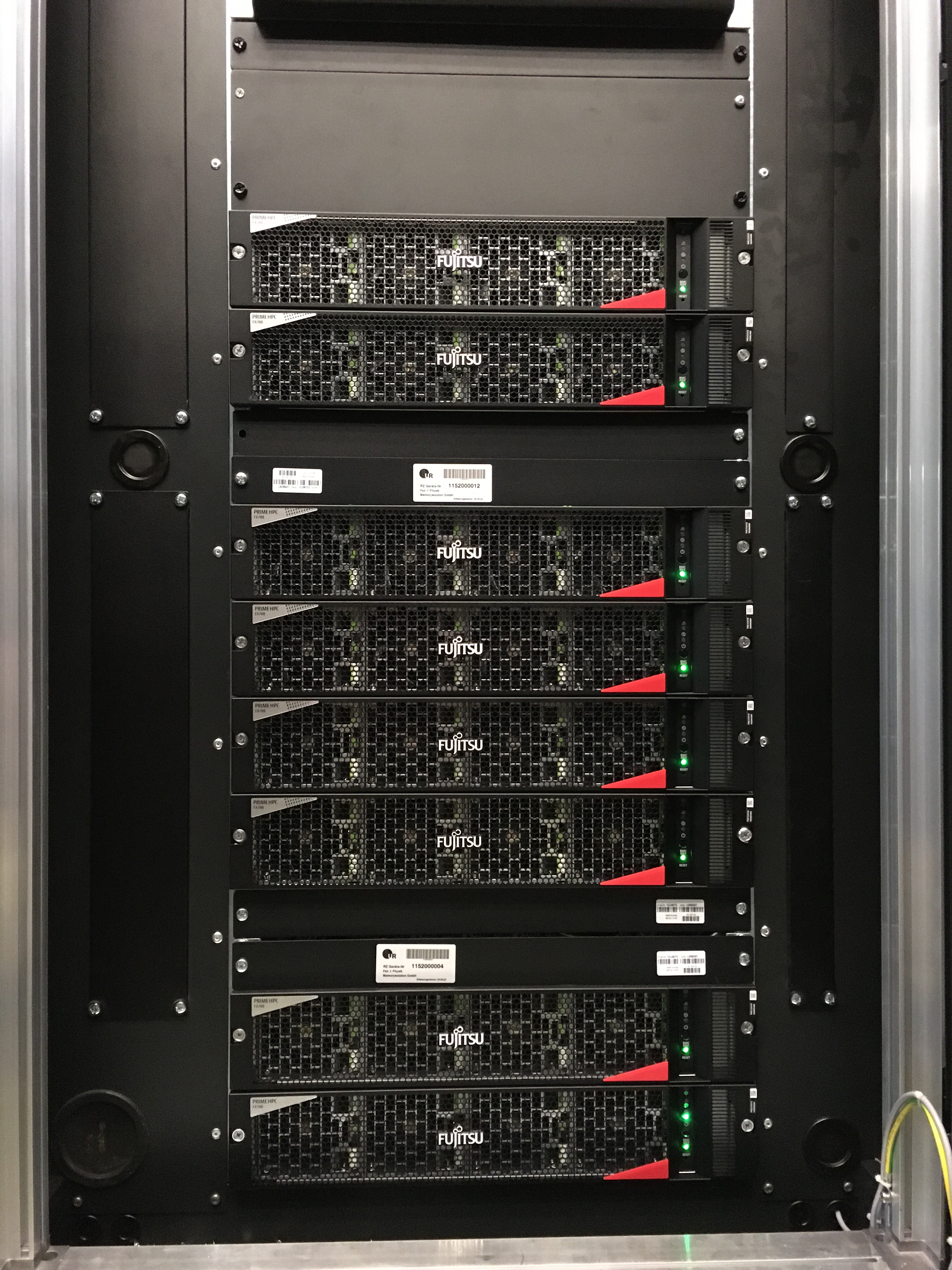}\\
            \caption{\label{fig:qp4}\qpfour cluster at the
            University of Regensburg.}
        \end{center}
\end{minipage}
\hfill
\begin{minipage}[t]{0.47\linewidth}
        \begin{center}
            \includegraphics[width=0.99\linewidth]{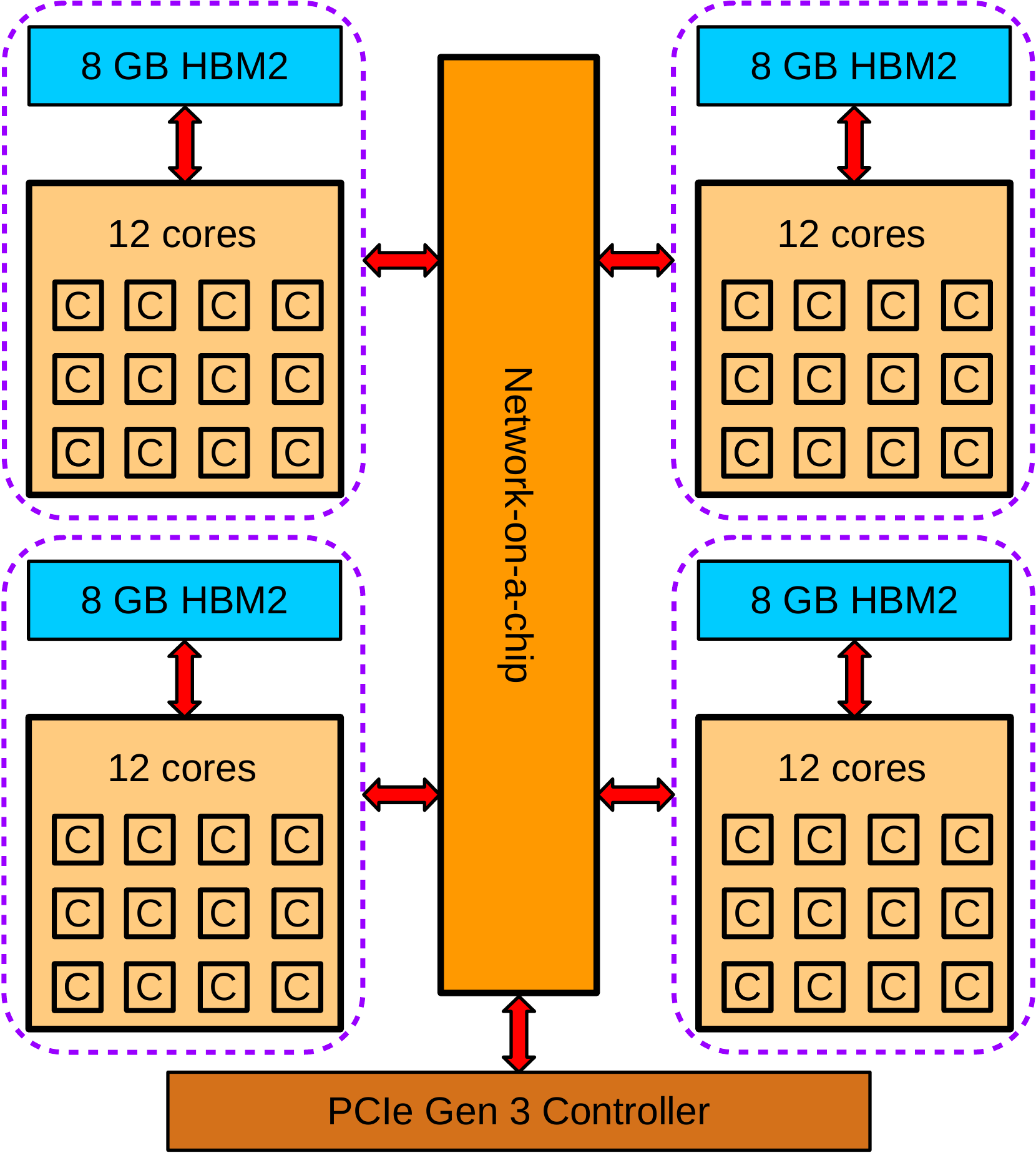}\\
            \caption{\label{fig:a64fx}High-level architecture of the Fujitsu
            A64FX CPU.}
        \end{center}
\end{minipage}
\end{figure}

\section{\qpfour}

\qpfour is the latest member of the QCD PArallel Computing Engine (QPACE) series.
The small-scale cluster features compute nodes from the Fujitsu PRIMEHPC
FX700 series.  \qpfour was deployed at the University of Regensburg in June 2020
and is shown in Fig.~\ref{fig:qp4}. The key characteristics of \qpfour are
\begin{itemize}\itemsep0mm
  \item 64 Fujitsu A64FX CPUs, 48 cores each (up to 4 NUMA domains), 1.8 GHz
    (see Fig.~\ref{fig:a64fx})
  \item 512-bit ARM Scalable Vector Extension (SVE)
  \item 177/354 TFlop/s peak in double/single precision (DP/SP)
  \item 2048 GB HBM2 memory total
  \item InfiniBand EDR interconnect (100 Gbit/s)
\end{itemize}
We run an open-source software stack using CentOS Stream 8, GCC 10.1 and
OpenMPI 4.0.  For storage we use the GlusterFS parallel file system.
For Lattice QCD we use Grid \cite{Boyle:2015tjk} and the Grid Python Toolkit
(GPT) \cite{Lehner:gpt}.

\section{Main memory throughput}

The nominal aggregate peak throughput of the HBM2 memories of a single node is
1 TB/s.  The de-facto standard benchmark for evaluation of memory throughput
is the STREAM benchmark \cite{stream}.  STREAM consists of four mini-benchmarks
(copy, scale, add, and triad).  The mini-benchmarks differ in computation and
in the load/store ratio of array elements.  For copy and scale the load/store
ratio is 1:1 per element, and for add and triad this ratio is 2:1.

\begin{figure}
\begin{minipage}[t]{0.49\linewidth}
        \begin{center}
            \includegraphics[width=0.99\linewidth]{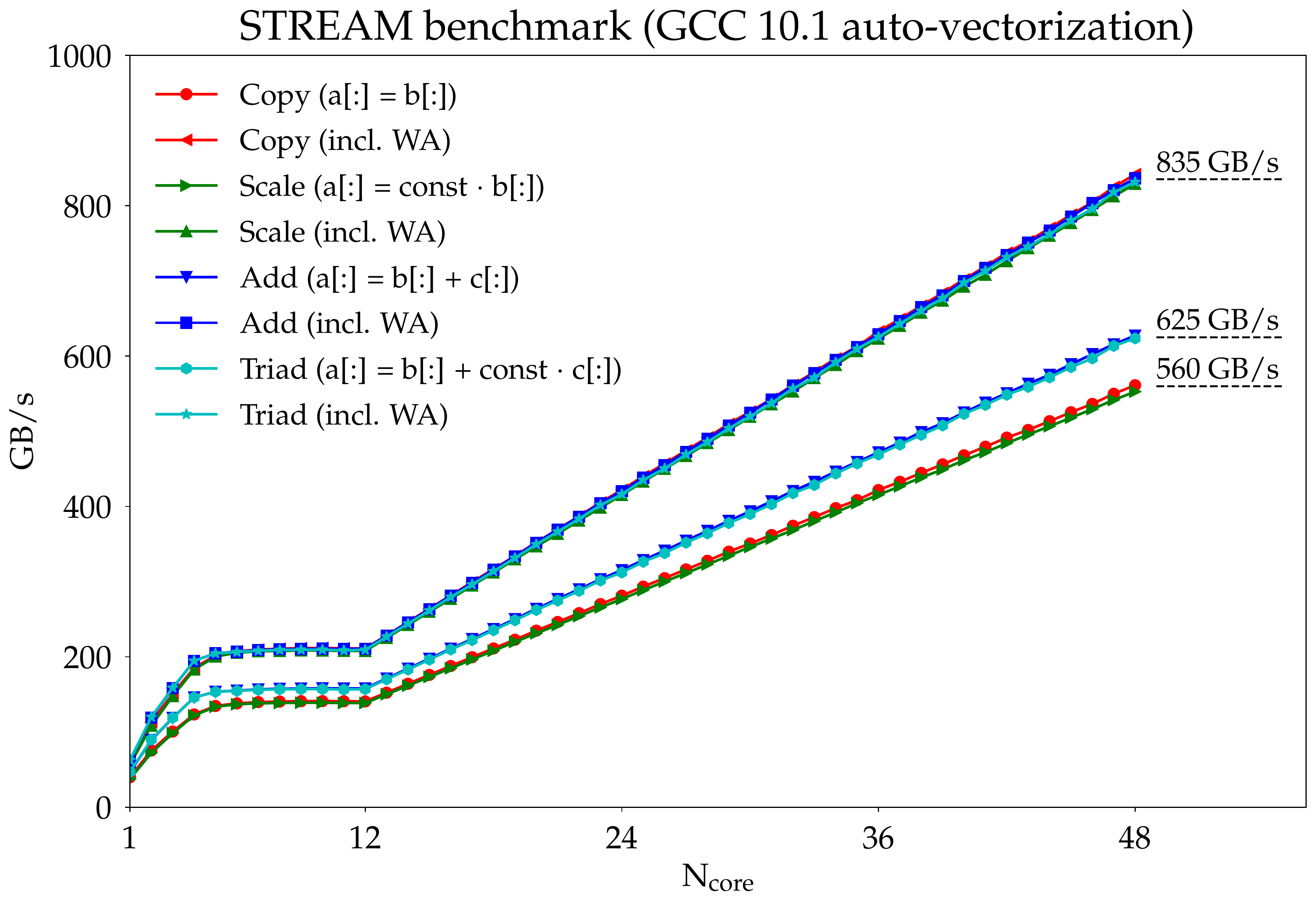}\\
            \caption{\label{fig:stream}STREAM benchmark.}
        \end{center}
\end{minipage}
\hfill
\begin{minipage}[t]{0.49\linewidth}
        \begin{center}
            \includegraphics[width=0.99\linewidth]{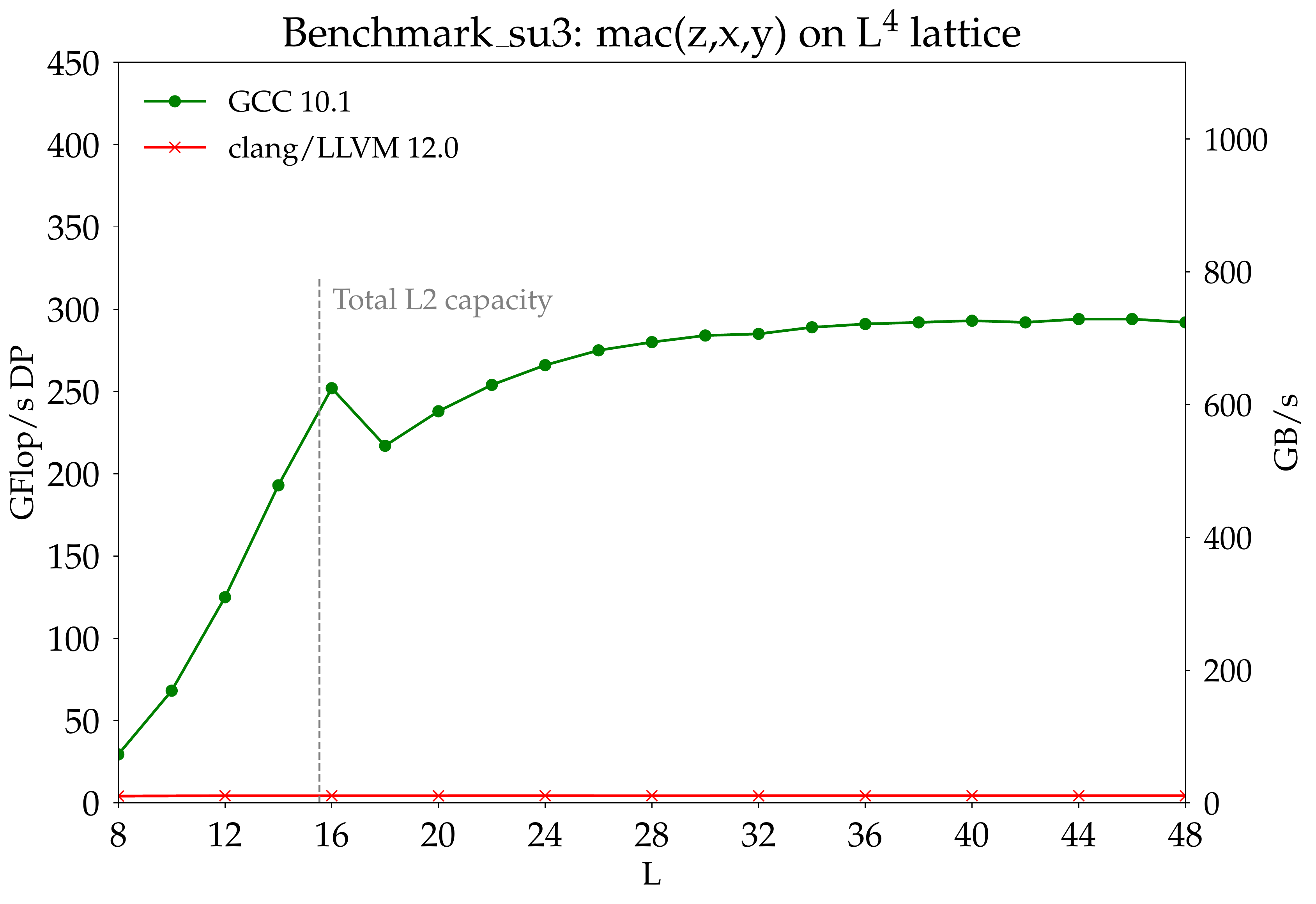}\\
            \caption{\label{fig:su3}SU(3) matrix
              multiplication. Arithmetic performance is proportional
              to data throughput.}
        \end{center}
\end{minipage}
\end{figure}

We show thread scaling of the STREAM benchmark on a single \qpfour node in
Fig.~\ref{fig:stream}.  Data throughput scales with the number of cores in use
if $\text{N}_\text{core} > 12$.  We measure up to
625 GB/s of benchmark data throughput.  The caches of the A64FX implement a
write-back policy, i.e., a cache block is loaded from main memory on a
write miss.  This is called Write Allocation (WA).  WA causes extra data traffic
and thus reduces the effective memory throughput.  Including WA traffic,
the data throughput is up to 835 GB/s.  We note that STREAM on a single
Fugaku node (48 compute cores, 2.2 GHz) yields comparable results despite the
difference in clock frequencies.

\section{Grid on \qpfour}

\subsection{Port to A64FX (512-bit SVE)}

\begin{figure}
\begin{minipage}[t]{\linewidth}
        \begin{center}
            \includegraphics[width=\linewidth]{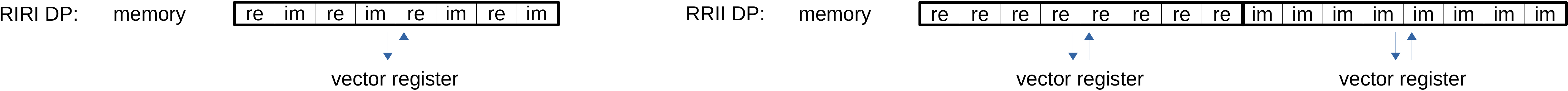}\\
            \caption{\label{fig:layout}Data layout of complex numbers
            in DP using 512-bit vector registers:
            alternating real and imaginary parts
            (RIRI) and separate real and imaginary parts (RRII).}
        \end{center}
\end{minipage}
\end{figure}

The ARM C Language Extensions (ACLE) provide access to SVE vector types and
SVE instructions in C/C++ \cite{Arm:ACLE}.  We use ACLE to implement Grid's
lower-level functions.  Grid stores complex numbers in memory alternating real
and imaginary parts (see Fig.~\ref{fig:layout}).  We refer to this layout as
RIRI.  We use hardware support for processing complex
numbers.  We published the details of the implementation in
\cite{Meyer:cluster18, Meyer:lattice18, Meyer:aplat20}.  Our Grid port
to the A64FX is available in upstream Grid
(\texttt{configure -{}-enable-simd=A64FX}).

Key computational kernels such as Wilson Dslash and the
performance-relevant part of the Domain Wall operator,
which we refer to as Domain Wall kernel and discuss in detail in
\cite{Alappat:cpe21}, are
specialized to the A64FX using ACLE and the RIRI layout.  Manual instruction
scheduling facilitates instruction latency hiding and optimal use of the
register file.  We use software
prefetching to optimize the data flow between compute cores and memory
hierarchy.
Specialized kernels are enabled by the
command line argument \texttt{-{}-dslash-asm} when starting binaries.

\subsection{Benchmarks}

Grid features a rich set of tests and performance benchmarks.  Here we present
a relevant subset of benchmarks to illustrate the performance of Grid on
\qpfour.

Benchmark\_su3 performs independent SU(3) matrix multiplication on each lattice
site in a 4d volume.  This benchmark is well suited for SVE compiler testing.
We show the performance of SU(3) matrix multiplication as a function of the
lattice volume on a single \qpfour node in Fig.~\ref{fig:su3}.
Using GCC 10.1 we achieve up to about 300 GFlop/s DP, which corresponds to a
benchmark data throughput of about 744 GB/s not taking into account WA traffic.
We also show the performance of the open source LLVM/clang SVE compiler.
LLVM/clang-based SVE compilers (including open source LLVM/clang, Arm's armclang,
and also Fujitsu's compiler in clang mode) introduce unnecessary
copy operations and therefore generate non-optimal code.  We did not test
Cray's LLVM/clang-based SVE compiler, but we expect similar results.
At present we recommend using GCC 10.1 or 10.2.

\begin{table}[b]
  \begin{minipage}[c]{0.47\linewidth}
    \begin{center}
      \small
      \begin{tabular}{lcccc}
        \toprule
        Volume & \multicolumn{2}{c}{1 MPI rank} & \multicolumn{2}{c}{4 MPI ranks}\\
               &  DP & SP  & DP  & SP \\
        \midrule
        $16^3\times 32$ & 360 & 893  & 321 & 619 \\
        $24^3\times 48$ & 350 & 990  & 383 & 827 \\
        $32^3\times 64$ & 339 & 1023 & 383 & 898 \\%\hline
        \bottomrule
      \end{tabular}
      \captionof{table}{\label{tab:wilson}Performance of the Wilson
        Dslash kernel on a single node in GFlop/s.}
    \end{center}
  \end{minipage}%
  \hfill%
  \begin{minipage}[c]{0.47\linewidth}
    \begin{center}
      \small
      \begin{tabular}{lcccc}
        \toprule
        Volume & \multicolumn{2}{c}{1 MPI rank} & \multicolumn{2}{c}{4 MPI ranks}\\
               &  DP  & SP  & DP  & SP \\
        \midrule
        $16^3\times 32\times 8$  & 484 & 949 & 408 & 803  \\
        $16^3\times 32\times 16$ & 477 & 958 & 409 & 815 \\
        $24^3\times 48\times  8$ & 476 & 960 & 432 & 875 \\
        \bottomrule
      \end{tabular}
      \captionof{table}{\label{tab:dwf}Performance of the Domain Wall kernel on a
        single node in GFlop/s.}
    \end{center}
  \end{minipage}
\end{table}

The performance of the Wilson Dslash kernel and the Domain Wall kernel
is shown in Table~\ref{tab:wilson} and \ref{tab:dwf}, respectively, on a
single \qpfour node.  We observe a mild performance penalty using 4 MPI
ranks compared to 1 MPI rank.  We attribute this penalty to the overhead
associated with inter-process communication using shared memory.
SP performance is nominally expected to be about twice the DP performance due to
the doubling of the Flop throughput.
We observe this for the Domain Wall kernel, while for the Wilson Dslash kernel
DP performance is lower than expected.  To explain this, we note that (a) Grid
uses different data layouts for SP and DP and (b) the Domain Wall kernel has
higher cache reuse than the Wilson Dslash kernel.
Performance variations for different lattice volumes are mild.
We attribute the variations to the details of cache reuse, which differ amongst
lattice volumes.  We note that GCC 10.1 and 10.2 deliver best performance.
LLVM/clang-based compilers and other versions of GCC perform worse.

\begin{figure}
\begin{minipage}[t]{0.49\linewidth}
        \begin{center}
            \includegraphics[width=0.99\linewidth]{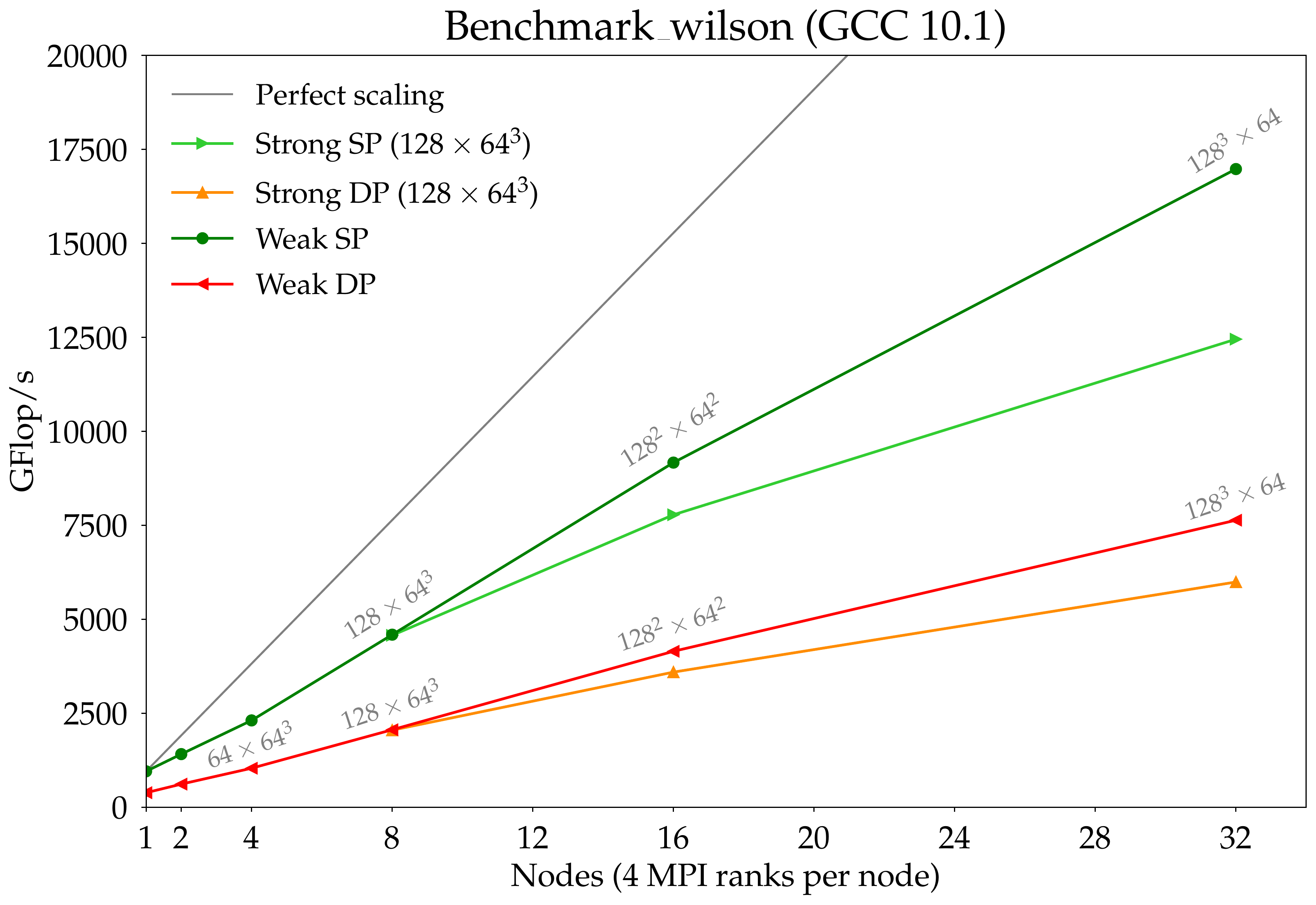}\\
            \caption{\label{fig:wilsonmpi}MPI scaling of the Wilson Dslash
            kernel.}
        \end{center}
\end{minipage}
\hfill
\begin{minipage}[t]{0.49\linewidth}
        \begin{center}
            \includegraphics[width=0.99\linewidth]{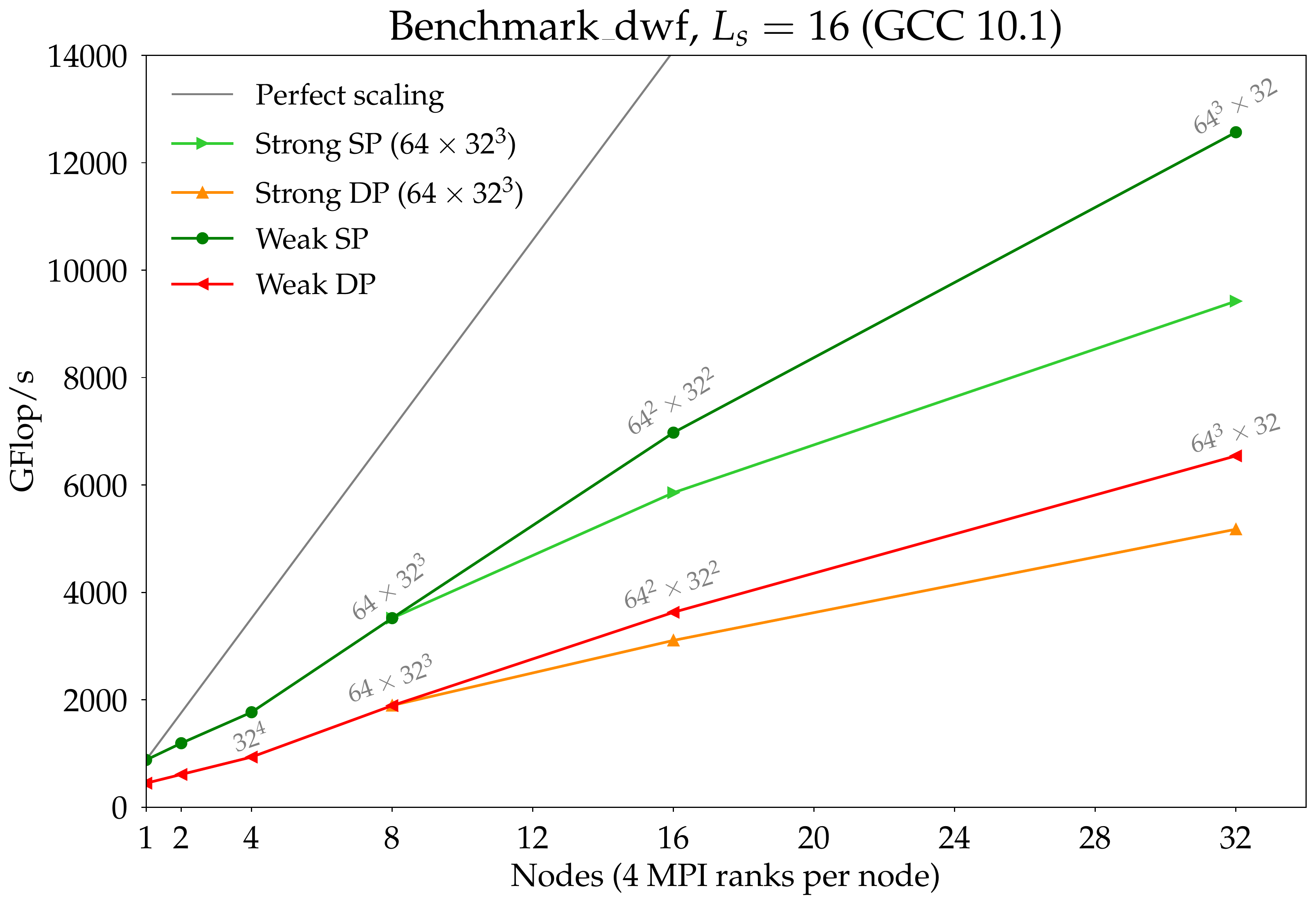}\\
            \caption{\label{fig:dwfmpi}MPI scaling of the Domain Wall
            kernel.}
        \end{center}
\end{minipage}
\end{figure}

The multi-node performance of key computational kernels using MPI is shown in
Figs.~\ref{fig:wilsonmpi} and~\ref{fig:dwfmpi}.
\qpfour has two separate InfiniBand partitions, and the maximum number of nodes
within each partition is 32.  Grid's MPI code path is non-trivial.
For instance, it includes many branches
in the instruction stream, multiple function calls, irregular memory access,
and communication over the network.
We observe a significant performance drop running
on multiple nodes compared to a single node despite large local volumes.
Differences between 1 rank per node and 4 ranks per node are marginal (< 3\%).
Further investigation is necessary to identify and resolve
bottlenecks in the MPI code path.

\subsection{Alternative layout of complex numbers}

Grid only supports the RIRI layout for complex numbers.  In a collaboration with
University of Erlangen we studied the Domain Wall kernel using an alternative data
layout for complex numbers on Fugaku \cite{Alappat:cpe21}.  We illustrate the
RRII layout in Fig.~\ref{fig:layout}.  Key characteristics of the RIRI and RRII
layout are shown in Table~\ref{tab:layoutchar}.  For RIRI we use the
\texttt{svcadd} ($z_1 \pm i z_2$) and \texttt{svcmla} ($z_1 + z_2\times z_3$)
intrinsics for processing complex numbers ($z_i\in \mathbb{C}$).
The corresponding instructions have high latencies and limitations on
pipeline usage (see \cite{Alappat:cpe21, Fujitsu:microarch} for details).
SVE does not support complex multiplication without addition of a third operand.
For RRII only real arithmetics is necessary, which comes with lower instruction
latencies and no limitations on pipeline usage.

\begin{figure}
        \begin{center}
            \includegraphics[width=0.5\linewidth]{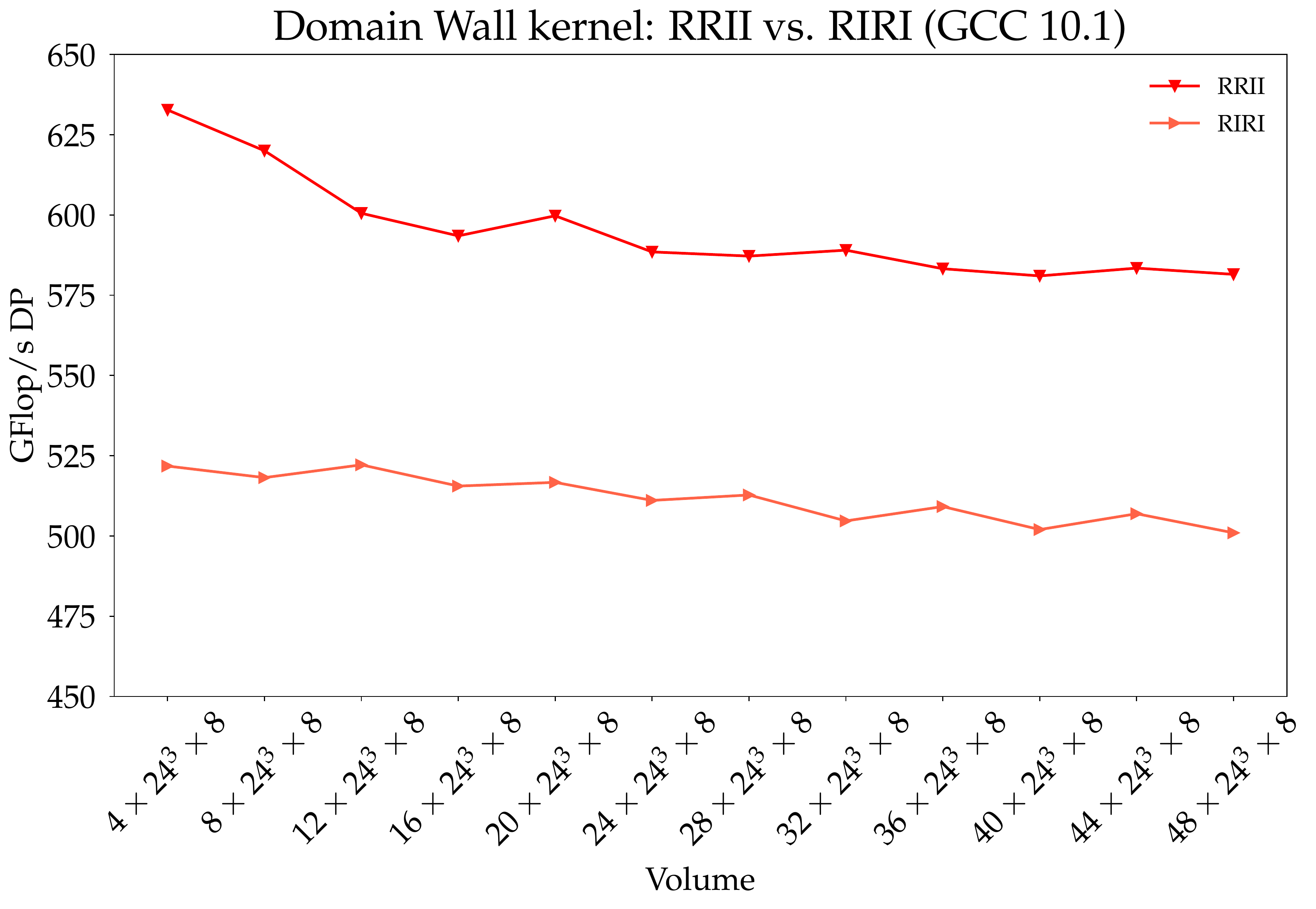}
            \caption{\label{fig:dwfrrii}Performance of the Domain
            Wall kernel using RRII and RIRI layout.}
        \end{center}
\end{figure}

We tested the RRII layout on the A64FX extending GridBench
\cite{Meyer:gridbench}, which features a subset of Grid.  We implemented the
Domain Wall kernel using ACLE and used datasets generated by Grid.
GridBench does not support MPI, therefore benchmarks
are restricted to a single node.  On Fugaku (48 cores, 2.2 GHz) the RRII
layout outperforms RIRI by up to 12\% \cite{Alappat:cpe21}.
On a single \qpfour node the RRII layout outperforms RIRI by up to about 20\%,
see Fig.~\ref{fig:dwfrrii}.
Instructions processing complex numbers predominantly use the FLA pipeline,
thereby leading to imbalanced pipeline usage.  Pipeline usage is balanced
using real arithmetics, which is the case for the RRII layout.
RIRI consumes up to 20\% more energy than RRII on Fugaku.
RRII energy consumption benefits further from shutting down one of the two
floating-point pipelines with minor impact on performance.
Currently we are not able to shut down pipelines on \qpfour due to
missing software support.  However, we expect energy savings comparable to, or
even higher than, Fugaku and comparable impact on performance.  We plan to
extend Grid to support the RRII layout in the future.

\begin{table}
\begin{center}
  \small
  \begin{tabular}{lll}
    \toprule
    & RIRI        & RRII \\
    \midrule
    Parallel lattice site updates     & 4 DP, 8 SP & 8 DP, 16 SP \\
    Full spinor projection to half spinor & \texttt{svcadd} ($z_1 \pm i z_2$) if applicable & Real arithmetics \\
    Reconstruction of full spinor     & \texttt{svcadd} ($z_1 \pm i z_2$) if applicable & Real arithmetics \\
    SU(3) $\times$ half spinor        & \texttt{svcmla} ($z_1 + z_2\times z_3$)       & Real arithmetics \\
    Real Flop per lattice site update & 1416 (need to compute $0 + z_2\times z_3$)       & 1320 \\
    FLA / FLB pipeline usage          & Imbalanced                            & Balanced \\
  \bottomrule
  \end{tabular}
\end{center}
\caption{\label{tab:layoutchar}RIRI and RRII layout characteristics for the
Domain Wall kernel ($z_i\in \mathbb{C}$).}
\end{table}

\section{Summary and outlook}

\qpfour comprises 64 Fujitsu A64FX CPUs interconnected by InfiniBand EDR.
The A64FX achieves outstanding performance for Lattice QCD applications.
GCC 10.1 and 10.2 achieve best overall performance, while other compilers
underperform.  An alternative layout of
complex numbers is beneficial on the A64FX, but has yet to be integrated into
Grid.  The necessary code modifications remain future work.

\section*{Acknowledgment}

We acknowledge funding of the \qpfour project provided by the Deutsche
Forschungsgemeinschaft (DFG) in the framework of SFB/TRR-55.  Furthermore, we
acknowledge support from the HPC tools team at Arm.  We thank Yasumichi Aoki,
Peter Boyle, Issaku Kanamori, and Yoshifumi Nakamura for valuable
discussions.

\bibliographystyle{JHEP_lat21}
\bibliography{references}

\end{document}